\documentclass[aps,showpacs,twocolumn]{revtex4}
\usepackage{graphicx}
\usepackage{amsmath}
\usepackage{amsfonts}
\usepackage{amssymb}
\usepackage{bm}
\usepackage{psfrag}
\begin{document}

\preprint{not yet sub}

\title{Chaotic filtering of moving atoms in pulsed optical lattices}

\author{T. Jonckheere, M. R. Isherwood and T. S. Monteiro}

\affiliation{Department of Physics and Astronomy, University College London,
Gower Street, London WC1E 6BT, U.K.}

\date{\today}

\begin{abstract}
We propose a mechanism for a velocity-selective device, which exploits the fundamental phenomenon
of dynamical localization. It would allow packets of cold atoms travelling through a pulsed optical
lattice in one direction to pass undisturbed, while dispersing atoms
travelling in the opposite direction. This mechanism is based on a chaotic diffusion process, with
a momentum-dependent diffusion coefficient. We analyse this classical effect, and we show that 
it is  {\it enhanced} by dynamical localization.
\end{abstract}


\maketitle
There is much current interest in the development of new techniques to manipulate
cold atoms. Recent work in atom optics has resulted in new devices termed 
 ``atom chips'' \cite{Hinds,Reichel} where cold atoms may be trapped and guided by
fields above a solid substrate. One might even in future envisage a 
multi-component chip combining arrays of micro-traps and atom wires, forming
a sort of atom optics version of an integrated circuit. Within such an atom chip,
techniques to, for example, split, transport and otherwise control the traffic
of atoms, might play an important role.

Optical lattices in particular are receiving much current attention
 in atom optics, in a wide range of experimental studies.
For instance, cold atoms in optical lattices have become a paradigm in the study of
quantum chaos; experiments on sodium and cesium atoms \cite{Raizen}
provided a textbook demonstration of Dynamical Localization (DL)
\cite{Casati,Fish}, the quantum suppression of classical chaotic diffusion.
In these experiments the atoms experience a periodically pulsed or driven standing
wave of light.

 The corresponding classical motion is fully 
chaotic for sufficiently strong driving. This implies that the energy of the system grows diffusively
with each consecutive kick: in the absence of phase-space barriers,
which are present only in the regular regime, the average energy $\langle p^2 \rangle$
of the particles is unbounded and this diffusive increase in energy
 continues indefinitely. It is characterised by a diffusion rate $D_0$, ie
$\langle p^2 \rangle =D_0 t$.
 However in the $quantum$ case this process is suppressed on the time-scale
of the so-called ``break-time'' $t^* \sim D_0/ \hbar^2$. After $t^*$ the atoms absorb no more
energy.

The series of ground-breaking experiments in \cite{Raizen}
 was followed by other  experiments with optical lattices
 probing a wide range
of quantum chaos phenomena including 
dynamical tunnelling 
\cite{Raiz3,NIST}, the effect of quantum loss of coherence on
dynamical localization \cite{Raiz2} and quantum accelerator modes 
\cite{Darcy}.
The experiments in \cite{NIST} employed BECs in optical lattices.
The dynamics of Bose-Einstein Condensates in optical lattices has
 recently become
of very broad interest \cite{Mott}. 

Here we show that a straightforward modification of the quantum chaos 
experiments can form
the basis for a device to control the traffic of cold atoms moving along
a channel in, say, an atom chip by selecting a specified velocity.
 It was recently
proposed \cite{Mon1} that atoms in double-well lattices, pulsed with unequal 
periods, will have diffusion rates which are momentum dependent: $D(p)= D_0 + 
C(p)$ where
$C(p)$ is a momentum dependent correction.
This can form the basis for a type of  ratchet dynamics: for a system with  
zero initial momentum, 
 $\langle p(t=0) \rangle =0$, particles with $p>0$, say, absorbed energy 
at a different
rate from particles with $p<0$, leading to a non-zero net current.
 While for that system $D(p)$ is locally asymmetric around $p=0$, it has
no particular  symmetry over a larger scale and has complex oscillations 
with respect
 to $p$, with contributions from several terms. Hence it is not suitable
for the velocity selector we propose here.  Below we show that there is a 
system which has a diffusion coefficient of very simple
form, with a single odd-parity  correction term, ie $C(p)=-C(-p)$. For 
experimentally
accessible parameters we can then select a regime where, for particles moving 
in one direction
$D^+(p)=D_0+C(p) \sim 0$ which means
 they absorb little or no energy; while particles moving
in the opposite direction experience an enhanced diffusion rate $D^-(-p)> 
D_0$.
While in the classical case, this effect is  confined to a narrow 
parameter
range, we show here that for the equivalent
quantum system, this ``filtering effect'' is in fact 
substantially
 $enhanced$ by DL relative to the classical case, and remains effective 
over a wider
parameter range. We investigated DL in this anomalous diffusion regime and 
find
that it is associated with a local ``break-time''
  $t^*(p) \sim D(p)/\hbar^2$. The ratio of energy absorbed by particles moving in opposite
directions, classically is $\sim D^+(p)/D^-(p)$ for short times
and $\sim 1$ for large $t$ ; in the quantum case,
due to dynamical localization, the corresponding ratio is 
$[D^+(p)/D^-(p)]^2
\sim cst$ for $t>t^*$.
If implemented, this would represent an atom optics application of what to 
date
remains a much studied fundamental physics phenomenon.

\begin{figure}[ht]
\psfrag{x}{\bf (a)}
\psfrag{y}{\bf (b)}
\psfrag{aaa}{$K=2.0$, $\hbar=0.25$}
\psfrag{bbb}{$K=3.2$, $\hbar=1.0$}
\psfrag{po}{$p$}
\includegraphics[height=2.5in,width=3.in]{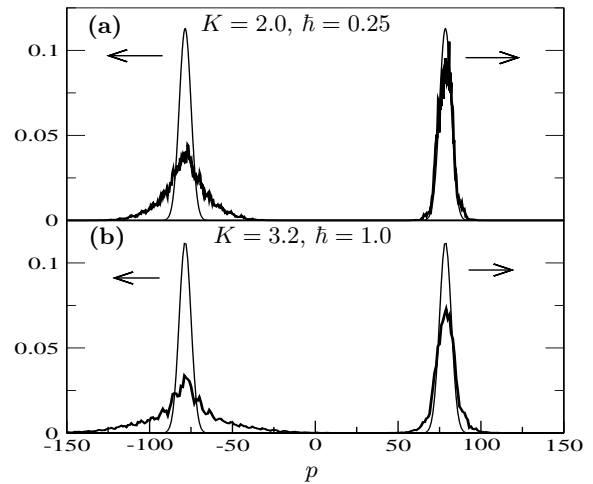}
\caption {Demonstration of the filtering effect: 
figure compares the effect of 50 pairs of pulses on 
two quantum wavepackets moving in
opposite directions. Initial wavepackets are shown with a thin line, while the final wavepackets
are shown with a bold line. Parameters : (a) for  kick strength $K=2.0$, $\hbar=0.25$, 
(b) for $K=3.2$, $\hbar=1.0$.
The values of $A$ and $b$ are, in both cases, $\pi/2$ and $0.01$ respectively. The initial wavepackets
are centred around $\pm p_0$, with $4 b p_0 = \pi$. 
The initial widths of the wavepackets are similar to the experimental values.
 The figure shows that while the right-moving wavepacket
is only slightly perturbed, the left-moving one is strongly dispersed.
 }
\label{Fig.1}
\end{figure}

In the dynamical localization experiment of \cite{Raizen,Raiz2,Raiz3},
the dynamics  is approximately given by the kicked-particle Hamiltonian :
 $H=\frac{p^2}{2} -  K \cos x \sum_n \delta(t-nT)$
where $K$ is the kick strength. The classical dynamics is obtained by
iterating the well-known ``Standard Map'':
$x_{i+1}=x_i + p_i T$; $ p_{i+1}= p_i + K \sin x_{i+1}$.
In the Standard Map, we can take $T=1$, without loss of generality, but
for the proposed system, we use a repeating cycle of unequally spaced kicks.
For simplicity, we take a length-2 cycle, with the spacing between kicks alternating 
between  $T_1$ and $T_2$. The Hamiltonian is now given by :
\begin{eqnarray}
\label{eq:hamiltonian}
H = \frac{p^2}{2} + V(x) \sum_{n=0}^{\infty} \sum_{M=1}^{2} 
\delta(t-nT_{tot}+\sum_{i=1}^M T_i)
\end{eqnarray}
with $T_{tot} = T_1 + T_2$.
This means that the first kick after $t=0$ is at $t=T_1$, the second kick at $t=T_1+T_2$, the next
one at $t=T_1+(T_1+T_2)$ and so on. We will consider only a small deviation from equally spaced
kicks, which can be defined with the small parameter $b$, $T_1 = 1+b$, $T_2 = 1-b$.
The spatial symmetry is broken  by a  linear term which alternates in sign:
$V(x)= -(K \cos x + A x (-1)^j)$ where $j$ is the kick number.
This Hamiltonian leads to the map:
\begin{eqnarray*}
x_i = x_{i-1} + p_{i-1} (1+b)\\
p_i = p_{i-1} + K \sin x_i + A \\
x_{i+1} = x_i + p_i(1-b)\\
p_{i+1} = p_i + K \sin x_{i+1} - A\\
\label{eq:map}
\end{eqnarray*}
This classical map was investigated previously~\cite{Cheon} in
the regular regime.
Here we have investigated the properties of this map in the chaotic
regime.
We have also investigated the equivalent quantum behaviour.
Since the Hamiltonian involves only delta-kicks, we consider the usual time evolution
operator in a matrix representation using a plane-wave basis.
 We find that while single kicks couple different quasimomenta,
the combined time evolution operator for two kicks does not. Hence
it is most convenient to iterate  the two-kick time-evolution operator.
In the usual plane-wave basis $|n\rangle = \frac{1}{2 \pi} \exp{(i n x)}$
this takes the form of a matrix $\hat{U}(2)$ with elements:
\begin{multline}
\label{eq:Uoperator}
\langle n | U^q | l \rangle = e^{-i((1+b)(l+q)^2 \hbar)}\\
 \sum_j e^{-i(1-b)(j+qa)^2 \hbar}
           \, J_{l-j+ka}(\frac{K}{\hbar})J_{j-n-ka}(\frac{K}{\hbar})
\end{multline}
where $q$ is a quasimomentum, $ka=int(q-A)$ and $qa=q-A-ka$. We solve for the quantum
time evolution by simply iterating repeatedly $\psi(2)={\hat U} \psi(t=0)$.
In Fig.~\ref{Fig.1} we demonstrate the effect on two quantum 
wavepackets by iterating Eq.~(\ref{eq:Uoperator}) over 50 pairs of kicks for 
two different sets of parameters ($K=2.0$ and $\hbar=0.25$ for (a), $K=3.2$ and $\hbar=1.0$  for (b)). 
 We show that the device functions as a sort of Maxwell demon: 
while  packets moving to the right pass the pulsed lattice
 relatively unperturbed, those moving
to the left are strongly dispersed. Below we analyse this effect.

For the Standard Map, at the lowest level of approximation,
the classical momenta at consecutive kicks are uncorrelated and evolve in time as a
``random-walk''. This means that the average momentum of a large ensemble of particles is unchanged,
and that the average energy grows linearly: the average
kinetic energy $\langle p^2\rangle / 2$ grows by $K^2/4$ at each consecutive kick.
In the absence of phase-space barriers 
the energy is unbounded and this diffusive increase in energy
 continues indefinitely. It is characterised by a diffusion rate $D_0$, ie
$\langle p^2 \rangle =D_0 t$ so for uncorrelated momenta $D_0=K^2/2$. 
However, correlations between sequences
of consecutive kicks give important corrections to the diffusion coefficient. The 2-kick correlation
for example, originates from correlations between adjacent kicks but one, 
$\langle V'(x_i) V'(x_{i-2}) \rangle$, and gives a correction $-K^2 J_2(K)$ to the diffusion
coefficient. The diffusion coefficient, including the first corrections, is: 
$D_0 = K^2 (1/2 -J_2(K) - (J_1(K))^2 +...)$~\cite{Lich,Shep}.
 These corrections have even been measured experimentally
with cold cesium atoms in pulsed optical lattices \cite{Raizacc}.


For the  Hamiltonian in Eq.~(\ref{eq:hamiltonian}), the correlations 
 take a modified form.
The corrections they induce are now momentum dependent: 
for an ensemble of particle with initial momentum $p_0$, the average energy spread at time $t$ is given by
\begin{multline}
\label{eq:energyt}
 \langle \left( p - p_0 \right)^2 \rangle  =
 K^2 t \left(\frac{1}{2}- \frac{1}{2}[J_1(K(1+b))^2 + J_1(K(1-b))^2]\right) \\
      - K^2 \Phi(t) C(2,p_0) + ...
\end{multline}
Where
\begin{multline*}
C(2,p_0) =J_0(2 b K) \cos(2 p_0 b - A (1-b)) \\ \left[ J_2(K(1+b)) + J_2(K (1-b)) \right]
\end{multline*}
 and  
\begin{eqnarray*}
\Phi(t) = \frac{1-J_0(2bK)^{t-2}}{1-J_0(2 b K)^2}
\end{eqnarray*}
Higher order correlations (4-kick, etc.) involve more Bessel functions, and induce corrections that may become
significant over part of the parameter range, but eq.~(\ref{eq:energyt}) has the main features.
The first correction in equation~(\ref{eq:energyt}) is due to 3-kick correlation, and is very similar 
to the $J_1(K)^2$ correction for the standard map.
 The $C(2,p)$ term is the modified version of the 2-kick correlation term which for the standard
map took the form $-K^2 J_2(K)$. This term has several important properties. Firstly, it is $p$-dependent
with the $cos(2 p b - (1-b) A)$ term. This means that particles with different momenta will absorb
energy at different rates. This feature is the basis of the filtering effect shown above. 
Secondly, it has a non trivial time-dependence, given by the function $\Phi(t)$. 
Since $b K \ll 1$, one has $J_0(2 b K) \simeq 1 - (b K)^2$. For $t \ll 1/ (b K)^2$,
 $\Phi(t)$ has a linear behaviour ($\sim t/2$), and the $C(2,p)$ term appears as a correction
to the diffusion coefficient, as in the standard map. However, for larger $t$, $\Phi(t)$ saturates to
the value $(b K)^2/2$, and the 2-kick correlation does not modify the energy growth  anymore.

All the expressions simplify considerably if we consider that $b \sim 0.01$
is a small deviation from period-one pulses, so
 $bA, bK <<1$. For  times $t<< 1/(b K)^2$,
we can write:
\begin{multline}
\label{eq:energytsimple}
   \langle \left( p - p_0 \right)^2 \rangle = K^2 t \; \bigg[\frac{1}{2}- 
J_1(K)^2 - \\J_2(K)\, cos(2 p_0 b-A)\bigg]
\end{multline}
Hence we have a local diffusion coefficient of the form
$D(p)\simeq D_0 - C(2,p)$, where $D_0 \simeq K^2 [1/2 - J_1(K)^2]$.

\begin{figure}
\psfrag{K}{$K$}
\psfrag{ppp}{$\langle (p-p_0)^2 \rangle - D_0 t$}
\psfrag{simple analytical formula  }{simple formula~(\ref{eq:energytsimple})}
\psfrag{full analytic formula}{full formula~(\ref{eq:energyt})}
\psfrag{numerical results}{numerical results}
\centerline{\includegraphics[height=8.cm,angle=-90]{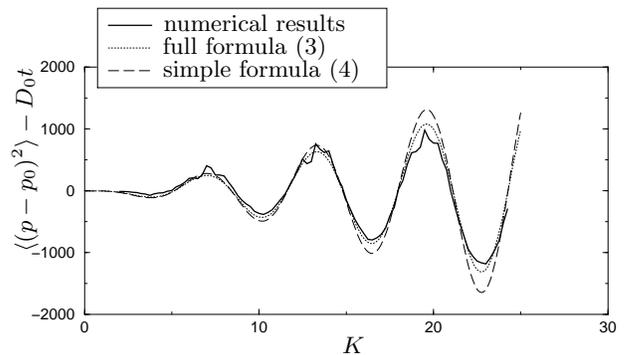}}
\caption{Average energy spread as a function of $K$, for an ensemble of 400,000 particles with a narrow
initial momentum distribution around $p_0 = \pi / (4 b)$, with $b=0.005$ and $A=\pi/2$. The $D_0 t$ term has
been removed (see text). The numerical results
are compared to the simple formula~(\ref{eq:energytsimple}), valid for $t \ll 1/(b K)^2$, and to the
full analytical formula~(\ref{eq:energyt}). }
\label{fig:EofK}
\end{figure}

Figure~\ref{fig:EofK} shows a comparison of this formula to numerical results:
the average energy spread after 20 kicks is plotted as a function of $K$, for 
an ensemble of 400,000 particles with a narrow initial momentum distribution around $ p_0 = \pi / (4 b)$,  with
$b=0.005$ and $A=\pi/2$. For these values, one has $\cos(2 p_0 b - A)=1$. For clarity, we have removed
the $p$-independent contribution $D_0$.
 We see that the numerical results agree 
very well with the formula~(\ref{eq:energytsimple}), showing the $J_2(K)$ oscillations. For the
larger values of $K$ shown on the figure, the effect of the function $\Phi(t)$ begins to be noticeable,
and formula~(\ref{eq:energytsimple}) over-estimates the result. The full analytic formula~(\ref{eq:energyt})
is also shown, with an excellent agreement over the whole range of $K$.

\begin{figure}
\psfrag{pmpo}{$\langle (p-p_0)^2 \rangle$}
\psfrag{x}{\bf (a)}
\psfrag{y}{\bf (b)}
\psfrag{z}{\bf (c)}
\psfrag{po}{$p_0$}
\psfrag{nk}{$n_{\mathrm{kick}}$}
\centerline{\includegraphics[height=7.0cm, angle=-90]{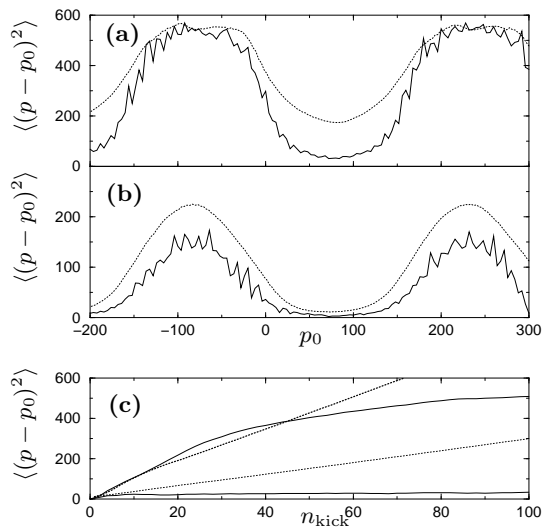}}
\caption{(a)-(b): average energy spread as a function of initial momentum, in the classical case 
(dotted curve) and the quantum case (full curve). The parameters are: (a) $K=3.2$, $\hbar=1$, 60 kicks for
the classical curve, 200 kicks for the quantum curve; (b) $K=1.7$, $\hbar=0.25$, 100 kicks for
the classical curve, 100 kicks for the quantum curve. In both case, $A=\pi/2$ and $b=0.01$. In (a), the number 
of kicks in the classical case has been chosen to obtain the same maximum value for the energy spread as
in the quantum case. The results show clearly that the 
diffusion coefficient oscillates as a function of $p$, with period $\pi / b$, and that the minimum of the
diffusion coefficient is lowered in the quantum case compared to the classical one.\\
(c)  Average energy spread as a function of kick number, for two wavepackets: one initially centred around
$p_0 = \pi /(4 b)$ (the two lower curves), the other initially centred around $p_0 = -\pi /(4 b)$ 
(the two upper curves). The classical results are shown as dotted curves, while the quantum results
are shown as full curves. One can see that the quantum wavepacket with slow energy spread localizes
extremely quickly, while the other wavepacket takes a much longer time to localize.
Parameters: $K=3.2$, $b=0.01$, $A=\pi/2$ and $\hbar=1$ in the quantum case.
}
\label{fig:EofP}
\end{figure}

Figure~\ref{fig:EofP}, (a) and (b), shows the average energy spread  as a function of the
initial momentum $p_0$, for two different sets of parameters.
In each case both the classical and the quantum results are shown. For the classical case
(dotted line on the figure)
we obtain, as expected from eq.~(\ref{eq:energytsimple}), a cosine behaviour in $p_0$,  with
a period $\pi/ b$.
The quantum results (full curve) are similar to the classical one. In (a), the number of kicks has been chosen
to obtain the same maximum energy spreading for the quantum and the classical results; this shows
clearly that the quantum minimum of the energy spreading is much lower than the classical one.

 This quantum effect is due to dynamical localization: near a maximum or
a minimum of the $\cos(2 p_0 b -A)$ function, the classical diffusion coefficient
 is enhanced or reduced, respectively.
Since the break-time, which gives the time needed by the quantum system to localize, is proportional
to the diffusion coefficient, one sees that near a maximum of the diffusion coefficient, the quantum
system takes longer to localize, and thus the quantum wavepacket absorbs energy for a longer time.
On the other hand, near a minimum of the diffusion coefficient, the quantum system localizes more
quickly, and the quantum wavepacket absorbs energy for a shorter time. More quantitatively,
this means that the break time now depends on initial momentum: $t^*(p_0) \sim D(p_0)/\hbar^2$. 
As a consequence, if the ratio of the maximum and minimum classical diffusion coefficient (or energy spread) is 
of the order of $D_+/D_-$, then for the quantum system we expect a ratio for the energy spread
of the order of $(D_+/D_-)^2$. In fig.~\ref{fig:EofP} (a) ($K=3.4$) for example, 
the ratio  $D_+/D_-$ is roughly 
of the order of 3, while the ratio of the quantum energy spreads is roughly 9.
This enhancement due to dynamical localization is further illustrated on fig.~\ref{fig:EofP} (c), where
the energy of the two wavepackets (one initially centred around $p_0=\pi/(4 b)$ (minimum), the
other around $p_0=-\pi/(4 b)$ (maximum)) is plotted as function of the kick number, for the classical and the
quantum case, for $K=3.4$. One sees that the quantum wavepacket with the slowest energy spread
localizes very quickly (a few kicks are enough),
 while the other wavepacket takes a much longer time to localize.

In sum, the filtering device results from a combination of two physical processes. Firstly, 
a momentum-dependent diffusion coefficient of a particular simple form which enables one
to select the velocities of interest by adjusting the sign and magnitude of $b$ (in
the range $b \sim 0.05 - 0.01$ typically). Secondly dynamical localization, which enhances
the classical effect by localizing quickly (slowly) the quantum wavepacket which absorbs 
energy slowly (fast).  While this quantum filter works well over the range $K \sim 1.5 -3.5$ the
corresponding classical one is effective only for a restricted number of kicks and over a narrower
range of $K \sim 1.5 - 1.8$. The parameter ranges of $K$ and $\hbar$ ($\hbar \sim 0.25 - 1.0$) are
accessible with current experiments.

T.M. thanks Thomas Dittrich, Sergej Flach and
Holger Schanz for helpful discussions. M.I acknowledges an
EPSRC studentship.The work was supported by EPSRC grant GR/N19519.

\end{document}